\documentstyle[11pt,epsf]{article}
\newcommand{\postscript}[2]
   {\setlength{\epsfxsize}{#2\hsize}
   \centerline{\epsfbox{#1}}}
\def\theequation {\thesection.\arabic{equation}}
\makeatletter\@addtoreset {equation}{section}\makeatother

\begin{document}

\title{\bf Semi-stability of embedded solitons \\ 
in the general fifth-order KdV equation}

\author{Yu Tan$^\dagger$, \hspace{0.2cm} 
Jianke Yang$^\dagger$, \hspace{0.2cm}
Dmitry E. Pelinovsky$^{\dagger\dagger}$\\ \\
{\small $^{\dagger}$
Department of Mathematics and Statistics, 
University of Vermont, Burlington, VT 05401, USA}  \\ \\
{\small $^{\dagger\dagger}$ Department of Mathematics, 
McMaster University, Hamilton, Ontario, Canada, L8S 4K1} }
\date{} 
\maketitle

\begin{abstract}
Evolution of perturbed embedded solitons in the general 
Hamiltonian fifth-order Korteweg--de Vries (KdV) equation 
is studied. When an embedded soliton is perturbed, 
it sheds a one-directional continuous-wave radiation. 
It is shown that the radiation amplitude is not minimal 
in general. A dynamical equation for velocity of 
the perturbed embedded soliton is derived. 
This equation shows that a neutrally stable embedded soliton is 
in fact semi-stable. When the perturbation increases the momentum 
of the embedded soliton, the perturbed state approaches 
asymptotically the embedded soliton, while when 
the perturbation reduces the momentum of the embedded soliton, 
the perturbed state decays into radiation. 
Classes of initial conditions to induce
soliton decay or persistence are also determined. 
Our analytical results
are confirmed by direct numerical simulations of the 
fifth-order KdV equation. 
\end{abstract}

\section{Introduction}

Embedded solitons are solitary-wave solutions of nonlinear 
evolution equations which reside at discrete points inside 
the continuous spectrum of the linear wave system. 
The existence of such waves has been known in various 
physical systems such as the fifth-order Korteweg--de Vries (KdV)
equations \cite{olver,akylas,groves}, extended nonlinear 
Schr\"odinger equations \cite{buryak,fujioka}, coupled KdV
equations \cite{grimshaw}, second-harmonic-generation (SHG)
system \cite{yangprl99}, massive Thirring model 
\cite{thirring,movingES}, three-wave system \cite{threewave}, 
and many others \cite{champneysphysD01,yangathens01}. 
In \cite{yangprl99}, such waves were given the name 
``embedded solitons'', and their distinct semi-stability 
property was revealed on heuristic ground. 
This semi-stability means that when a perturbation increases 
a certain positive-definite quantity (energy or momentum) 
associated with the embedded soliton, then 
the perturbed state approaches asymptotically the embedded soliton. 
However, when the perturbation decreases energy (momentum) of 
the embedded soliton, the perturbed state decays into radiation. 

The semi-stability property was later proved rigorously 
for embedded solitons in the perturbed integrable fifth-order 
KdV equation \cite{yangstudy01} and in the generalized SHG system 
\cite{peliyang}. The method in \cite{yangstudy01} follows 
the soliton perturbation technique and describes embedded solitons 
as critical points of a first-order dynamical system. Both 
location and stability of critical points can be studied within 
the reduced system. The other paper \cite{peliyang} develops 
the normal form analysis which relies on the known existence 
and linearized stability properties of embedded solitons. It proves 
the nonlinear semi-stability of embedded solitons 
through wave resonance mechanisms.
The latter method does not rely on the integrability of 
the original system, and it can be extended to any 
embedded-soliton-bearing system under certain assumptions.

Nonlinear semistability of embedded solitons is an interesting 
phenomenon because it occurs beyond the linear stability. 
The linear stability of solitary waves in the fifth-order 
KdV equation was studied by using the energy-momentum methods 
\cite{DK,Levandovsky} and the symplectic Evans matrix methods 
\cite{bridges}. A solitary wave is linearly unstable if
the linearization operator possesses an eigenvalue with a positive
real part. 
If all eigenvalues lie on the imaginary axis, the wave 
is called neutrally stable. A neutrally stable wave can however still be unstable 
due to algebraic instabilities \cite{PG2}.
For embedded solitons, the situation is different. 
Single-hump embedded solitons are generally neutrally 
stable in the linearized problem. 
However, a discrete zero eigenvalue of the linearization 
operator is embedded inside the continuous spectrum of this 
operator. Because of this, a nonlinear resonance between 
the embedded zero eigenvalue and the continuous spectrum can 
tunnel energy of a perturbed embedded soliton into continuous-wave 
radiation. The energy loss does not always destroy the 
embedded soliton though. If the perturbation increases the
energy (momentum) of the embedded soliton, then the radiation 
becomes weaker and weaker as the perturbed state asymptotically 
approaches the embedded soliton. But if the perturbation decreases 
the energy (momentum) of the embedded soliton, 
the radiation becomes stronger and stronger, and the embedded soliton
is then destroyed. Thus, the semi-stability of embedded solitons 
is an intrinsically nonlinear phenomenon beyond linear stability. 

Besides single-humped embedded soliton in the fifth-order KdV equation, 
multi-humped embedded solitons may also exist \cite{groves,champneys}. 
However, they are typically linearly unstable 
\cite{champneysphysD01,yangathens01}. 
The nonlinear semi-stability may occur in the system 
only if the linear instability is suppressed. 
Thus, it makes sense to consider here only neutrally stable 
single-humped embedded solitons. 

We emphasize that semi-stability is not the same as instability. 
By controlling the energy (momentum) of the initial perturbation, 
we can induce either asymptotic persistence of the embedded soliton 
or its rapid disappearance. This is an ideal mechanism for 
switching (quantization) applications. 

In this article, we study the evolution of perturbed neutrally stable 
embedded solitons in a general Hamiltonian fifth-order KdV equation. 
We use the normal form analysis developed in \cite{peliyang} 
but simplify many statements and proofs. We show that when 
an embedded soliton in the fifth-order KdV equation is perturbed, 
it sheds the continuous-wave radiation in front of the soliton. 
The radiation amplitude is not minimal in general. 
We also derive the velocity equation for the perturbed 
state which proves the semi-stability property of embedded 
solitons. In addition, we determine what initial condition
leads to soliton decay, and what initial condition leads to 
soliton persistence. 
Numerical simulations of the fifth-order KdV 
equation show excellent agreement with the analytical predictions. 

\section{Tail amplitudes of symmetric nonlocal waves}

We consider the general Hamiltonian fifth-order KdV equation
\begin{equation} \label{u}
u_t+u_{xxx}+u_{xxxxx}+[N(u)]_x=0, 
\end{equation}
where the nonlinear term $N(u)$ is of the form
\begin{equation} \label{N}
N(u)=\alpha_0 u^2+\alpha_1 uu_{xx}+\alpha_2u_x^2+\alpha_3u^3. 
\end{equation}
The equations above are Hamiltonian if and only if $\alpha_1=2\alpha_2$
\cite{yangstudy01}. The conserved Hamiltonian functional $H(u)$ is
\begin{equation}
H(u)=\int_{-\infty}^{\infty} \left[ \frac{\alpha_0}{3}u^3-\frac{1}{2}u_x^2
+\frac{1}{2}u_{xx}^2-\frac{\alpha_1}{2}uu_x^2+
\frac{\alpha_3}{4}u^4\right] dx. 
\end{equation}
When Eq. (\ref{u}) is Hamiltonian, it also conserves a positive-definite 
quantity
\begin{equation} \label{momentum}
P(u)=\int_{-\infty}^{\infty} u^2 dx, 
\end{equation}
which is interpreted as momentum. Whether Eq. (\ref{u}) is
Hamiltonian or not, it always conserves the quantity 
$M(u) = \int_{-\infty}^{\infty} u dx$, 
which is interpreted as mass. 

We look for moving stationary solutions in Eq. (\ref{u}) of the form
\begin{equation} \label{wave}
u(x, t)=U(x-c \hspace{0.05cm} t)\equiv U(\xi), 
\end{equation}
where $c$ is the wave velocity. Substituting Eq. (\ref{wave}) into 
(\ref{u}) and integrating once, 
we obtain the ordinary differential equation
(ODE) for $U(\xi)$ as
\begin{equation}\label{ode}
U_{\xi\xi\xi\xi}+U_{\xi\xi}-c \hspace{0.05cm} U+N(U)=0,
\end{equation}
subject to zero boundary conditions at infinity: 
$\lim_{|\xi| \to \infty} U(\xi) = 0$. When $U \to 0$, 
Eq. (\ref{ode}) becomes linear, and its characteristic equation is quartic. 
The roots of this characteristic equation are $\pm ik$ and $\pm \kappa$, where
\begin{equation} \label{k}
k=k(c)=\sqrt{\frac{\sqrt{1+4c}+1}{2}}, \;\;\;
\kappa=\kappa(c)=\sqrt{\frac{\sqrt{1+4c}-1}{2}}. 
\end{equation}
When $c>0$, roots $\pm ik$ are imaginary and $\pm \kappa$ real; 
when $0 > c > -1/4$, all four roots are imaginary; when 
$c < -1/4$, all four roots are complex. 
The embedded soliton, if it exists, arises from 
a saddle-node bifurcation \cite{champneys}, when the stable-unstable 
manifolds (real roots) correspond to the exponential tails of 
the embedded soliton, while the center manifold 
(imaginary roots) correspond to the tails of the continuous-wave 
radiation whose amplitude vanishes. Therefore, the embedded soliton 
may exist in Eq. (\ref{ode}) only for $c > 0$. 
We notice that non-embedded solitary waves with oscillatory and 
decaying tails may exist in Eq. (\ref{ode}) for $c < -1/4$ 
\cite{benilovmalomed} but such solutions are beyond 
the scope of this paper.

Only symmetric embedded solitons of Eq. (\ref{ode}) are considered here. 
Based on energy flux consideration, it is generally believed 
that asymmetric embedded solitons of Eq. (\ref{ode})
do not exist \cite{benilov,joshi}. If the system (\ref{u}) is
the perturbed integrable fifth-order KdV equation, 
the non-existence of asymmetric embedded solitons in Eq. (\ref{ode}) 
was proved in \cite{yangstudy01}. 

When $c > 0$, the tail amplitude of symmetric nonlocal waves 
of Eq. (\ref{ode}) is given asymptotically as
\begin{equation} \label{tail}
U(\xi; c, \delta) \longrightarrow r(c, \delta)
\sin(k|\xi|+\delta), \hspace{0.7cm}
|\xi| \rightarrow \infty. 
\end{equation}
Here $r$ is the amplitude of the oscillatory tail, 
and $\delta$ is the tail phase. 
Embedded solitons can be found numerically by a shooting method 
for Eq. (\ref{ode}) with $U'(0) = U'''(0) = 0$. 
Parameters of the shooting method $U(0)$ and $U''(0)$ satisfy 
only one condition that removes the exponentially growing term 
$\sim e^{\kappa |\xi|}$. It implies that one parameter in
symmetric nonlocal wave solutions of Eq. (\ref{ode}) is free 
in addition to the wave velocity $c$. We choose this free parameter 
to be the tail phase $\delta$. The tail amplitude $r(c,\delta)$ may 
vanish at certain discrete velocity values $c = c_{\rm ES}$ 
(in such cases, the phase $\delta$ becomes irrelevant). When it 
happens, we get embedded solitons. The $\mbox{sech}^2$ embedded 
solitons in Eq. (\ref{u}) has been investigated comprehensively 
in \cite{olver,yangstudy01}, and a family of one, two, and $\infty$ 
embedded solitons have been shown to exist in different parameter 
regions. Single-humped and multi-humped solutions of (\ref{ode}) 
were studied also in \cite{champneys}. We assume here the 
generic case of a co-dimension one bifurcation, when 
the embedded-soliton velocity $c_{\rm ES}$ is a simple zero 
of the tail amplitude $r(c,\delta)$, i.e., 
\begin{equation} \label{root}
r(c, \delta) = R(\delta) (c-c_{\rm ES}) + {\rm O}(c - c_{\rm ES})^2, 
\end{equation}
where 
\begin{equation}  \label{R}
R(\delta) = \frac{\partial r}{\partial c}(c_{\rm ES}, \delta) \,\ne 0
\end{equation}
is the slope of the tail amplitude $r$ of the symmetric nonlocal waves 
at the embedded-soliton velocity $c_{\rm ES}$ and phase $\delta$.
 
In the rest of this section, we derive the analytical 
expression for the tail amplitude $r(c,\delta)$ when the nonlocal 
wave velocity $c$ is close to the embedded-soliton
velocity $c_{\rm ES}$. For this purpose, we expand the nonlocal solution
$U(\xi; c, \delta)$ as a perturbation series:
\begin{equation} \label{expansion}
U(\xi; c, \delta) = U_{\rm ES}(\xi) + 
(c-c_{\rm ES}) U_1(\xi;\delta) + O(c-c_{\rm ES})^2. 
\end{equation}
When this expansion is substituted into Eq. (\ref{ode}), 
the function $U_1(\xi;\delta)$ is found to satisfy the following
inhomogeneous linear equation: 
\begin{equation} \label{inhomo}
{\cal L}U_1(\xi; \delta)=U_{\rm ES}(\xi), 
\end{equation}
where $\cal{L}$ is the linearization operator of Eq. (\ref{ode})
at embedded-soliton velocity $c_{\rm ES}$, i.e., 
\begin{equation} \label{L}
{\cal L}=\frac{d^4}{d\xi^4}+\frac{d^2}{d\xi^2}-c_{\rm ES}
+2\alpha_0 U_{\rm ES}+3\alpha_3 U_{\rm ES}^2+\alpha_1 
\frac{d^2 U_{\rm ES}}{d \xi^2} + 
\alpha_1 \frac{d}{d\xi} \left( U_{\rm ES}\frac{d}{d\xi}\right). 
\end{equation}
Here the Hamiltonian condition $\alpha_1=2\alpha_2$ has been 
utilized. Note that the operator ${\cal L}$ is self-adjoint 
in the Hamiltonian case. 

In order to solve the inhomogeneous equation (\ref{inhomo}), 
we need to know homogeneous solutions. Operator ${\cal L}$ 
has four homogeneous solutions $\psi_n \; (1\le n\le 4)$. 
The first solution is symmetric and bounded with asymptotic behavior:
\begin{equation}\label{psi1}
\psi_1(\xi) \longrightarrow \sin(k_r |\xi|+\delta_{\rm s}), \hspace{0.5cm}
|\xi| \rightarrow \infty,
\end{equation}
where $k_r \equiv k(c_{\rm ES})$ is the resonant wavenumber and 
$\delta_{\rm s}$ is the tail phase. The second solution is 
anti-symmetric and bounded with asymptotic behavior:
\begin{equation} \label{psi2}
\psi_2(\xi) \longrightarrow \sin(k_r \xi\pm \delta_{\rm a}), \hspace{0.5cm}
\xi \rightarrow \pm \infty,
\end{equation}
where $\delta_a$ is the tail phase. 
The third solution is anti-symmetric and localized, 
$\psi_3(\xi)= U_{\rm ES}'(\xi)$. 
It is related to spatial translational
invariance of the system (\ref{u}). The last solution 
$\psi_4(\xi)$ is symmetric and unbounded. 

Now we can solve the inhomogeneous equation (\ref{inhomo}) for $U_1$. 
Since we only consider symmetric nonlocal waves, the correction 
term $U_1(\xi)$ is symmetric as well. 
Note that $\partial U(\xi; c_{\rm ES}, \delta)/\partial c$ 
is an inhomogeneous solution of 
Eq. (\ref{inhomo}) for any $\delta$ values. This can be seen by 
differentiating Eq. (\ref{ode}) with respect to $c$, and then 
setting $c$ as $c_{\rm ES}$. 
Thus the general bounded symmetric solution of Eq. (\ref{inhomo})
can be written as
\begin{equation} \label{linearsolution}
U_1(\xi)=\frac{\partial U}{\partial c}(\xi; c_{\rm ES}, \delta_0)
+\gamma_1 \psi_1(\xi). 
\end{equation}
Here $\delta_0$ is any fixed phase, 
and $\gamma_1$ is an arbitrary constant. The homogeneous solutions 
$\psi_2(\xi)$ and $\psi_3(\xi)$ are not included as they are 
anti-symmetric, while the solution $\psi_4(\xi)$ is unbounded. 

The asymptotic oscillatory behavior of a general $U_1$ 
solution (\ref{linearsolution}) can be obtained 
from Eqs. (\ref{tail}) and (\ref{psi1}) as 
\begin{equation} \label{asym1}
U_1(\xi; \delta) \longrightarrow R(\delta_0)\sin(k_r|\xi|+
\delta_0) + \gamma_1 \sin(k_r|\xi|+\delta_{\rm s}), 
\hspace{0.5cm} |\xi| \rightarrow \infty. 
\end{equation}
where function $R(\delta)$ is 
defined by Eq. (\ref{R}). On the other hand, expanding 
the tail asymptotics (\ref{tail}) of the nonlocal wave 
$U(\xi; c, \delta)$ into a power series of $(c-c_{\rm ES})$ 
and then comparing it with Eq. (\ref{expansion}), we conclude 
that the asymptotic behavior of a general $U_1$ solution
should also be
\begin{equation} \label{asym2}
U_1(\xi) \longrightarrow R(\delta)\sin(k_r|\xi|+\delta), 
\hspace{0.5cm} |\xi| \rightarrow \infty.
\end{equation}
For convenience, we fix here $\delta_0$ as $\delta_0 = 
\delta_s + \pi/2$ and then equate the two asymptotics 
(\ref{asym1}) and (\ref{asym2}). As a result, we find that 
\begin{equation}
\label{RRRRR}
R(\delta)=\frac{R(\delta_{\rm s}+\frac{\pi}{2})}
{\sin(\delta-\delta_{\rm s})}, 
\end{equation}
and $\gamma_1=R(\delta_{\rm s}+\frac{\pi}{2})\cot(\delta-\delta_{\rm s})$. 
Consequently, to leading order in $(c-c_{\rm ES})$, 
the tail amplitude $r$ of symmetric nonlocal waves is
\begin{equation} \label{r}
r(c,\delta)=\frac{R(\delta_{\rm s}+\frac{\pi}{2})}
{\sin(\delta-\delta_{\rm s})} (c-c_{\rm ES}) + 
{\rm O}(c-c_{\rm ES})^2. 
\end{equation}
From this formula, we conclude that at a given velocity $c$ 
close to $c_{\rm ES}$, the tail amplitude $|r|$ of 
symmetric nonlocal waves is minimal at the value 
$\delta = \delta_{\rm min}$, where 
\begin{equation} \label{deltamin}
\delta_{\rm min} = 
\left( \delta_{\rm s}+\frac{\pi}{2} \right) \;\mbox{mod}\: (\pi).
\end{equation}
Formulae (\ref{r}) and (\ref{deltamin}) are the main results of this
section. 

\section{Dynamics of embedded solitons under perturbations}

In this section, we study dynamics of a linearly neutrally stable 
embedded soliton under small perturbations. 
We assume that the linearization operator has no unstable 
eigenvalues. This assumption is necessary as 
the weak nonlinear semi-stability would be ineffective in the
presence of strong linear instability. 
We also assume that the linearization operator has 
no discrete non-zero embedded eigenvalues. 
This assumption is necessary as the non-zero embedded 
eigenmodes may be in resonance with the continuous spectrum 
through nonlinear coupling, thus affecting the dynamics of 
embedded solitons. Thirdly, we assume that the zero 
eigenvalue corresponds to the localized eigenfunction 
$U_{\rm ES}'(\xi)$ and has algebraic multiplicity two
with an associated eigenfunction $\partial U/\partial c(\xi;\delta)$.
For embedded solitons, the zero eigenvalue is always embedded 
into the continuous spectrum of the linearization operator. 
Eigenvalues that correspond to localized eigenfunctions 
occur as zeros of the Evans function (a determinant of 
scattering coefficients) \cite{bridges}. The algebraic multiplicity 
of eigenvalues is defined as the multiplicity of zeros of 
the Evans function. Thus, our last assumption is that 
the Evans function has a double zero at $\lambda = 0$ of the 
linearized operator. 

We will use below the internal perturbation analysis described 
in \cite{peliyang} (see also \cite{PG2}). The idea is to 
recognize that under small perturbations, the eigenfunctions 
$U'_{ES}(\xi)$ and $\partial U/\partial c(\xi;\delta)$
for the double embedded eigenvalue $\lambda = 0$ of the 
linearized problem renormalize the location and velocity 
of the embedded soliton. 
For small perturbations, the velocity $c(t)$ of the embedded soliton 
changes on a slow time scale. 
We will derive a dynamical equation for 
$c(t)$ by separating the slow and fast changes in evolution 
of a perturbed embedded soliton. 

In the moving coordinate,
\begin{equation} \label{movingsystem}
\xi=x-\int_0^t c\: dt - x_0, 
\end{equation}
the fifth-order KdV equation (\ref{u}) can be written as 
\begin{equation}
u_t-c \hspace{0.05cm}
u_{\xi}+u_{\xi\xi\xi}+u_{\xi\xi\xi\xi\xi}+\left[N(u)\right]_{\xi}=0. 
\end{equation}
We expand the perturbed embedded soliton and its slowly 
varying velocity into the following perturbation series: 
\begin{equation}  \label{uexpand}
u(\xi, t)=U_{\rm ES}(\xi)+\epsilon c_1(T)u_1(\xi,t)+
\epsilon^2 u_2(\xi, t, T)+ {\rm O}(\epsilon^3), 
\end{equation}
and 
\begin{equation} \label{Vexpand}
c(T)=c_{\rm ES}+\epsilon c_1(T) + {\rm O}(\epsilon^2),  
\end{equation}
where $T=\epsilon t$, and $\epsilon$ is a small parameter. 
At order $\epsilon$, we obtain the governing 
equation for $u_1(\xi,t)$ as
\begin{equation} \label{u1}
u_{1t}+\left[{\cal L}u_1\right]_{\xi}= U_{\rm ES}'(\xi),
\end{equation}
where ${\cal L}$ is the same linearization operator as 
defined in Eq. (\ref{L}). 
The initial condition for Eq. (\ref{u1}) can be obtained from 
Eq. (\ref{uexpand}) as
\begin{equation}\label{initial}
u_1(\xi,0) = 
\frac{u(\xi,0)-U_{\rm ES}(\xi)}{\epsilon c_1(0)}.
\end{equation}
The initial value for the soliton velocity $c_1(0)$ can be found 
by projecting the initial deviation 
$u(\xi,0)-U_{\rm ES}(\xi)$ 
onto $\partial U(\xi;c_{\rm ES},\delta_{\rm a}) 
/ \partial c$, where the phase $\delta_{\rm a}$ is given by (\ref{psi2}). 
The projection is based on the spectral decomposition developed in 
Appendix A (see (\ref{decomposition})) and is given by 
\begin{equation}
\label{projection}
\epsilon c_1(0) = \frac{\int_{-\infty}^\infty U_{\rm ES}(\xi) 
\left[ u(\xi, 0) - U_{\rm ES}(\xi) \right] d\xi}{ 
\int_{-\infty}^{\infty} U_{\rm ES}(\xi) 
\frac{\partial U}{\partial c}(\xi;c_{\rm ES},\delta_{\rm a})d\xi}. 
\end{equation}

Next,  we solve the inhomogeneous equation (\ref{u1}). 
We adopt a less formal but more intuitive approach here. 
A more rigorous calculation of the same results
is presented in Appendix A.

The inhomogeneous term in Eq. (\ref{u1}) acts as
a driving localized force. The homogeneous part at 
large $|\xi|$ values supports oscillatory solutions 
with wavenumber $k_r$. Thus due to forcing and resonance, 
these oscillatory tails will be excited over time. 
The group velocity of these oscillatory tails in the 
moving frame (\ref{movingsystem})
can be found from the dispersion relation as 
$c_{\rm gr} = 2k_r^2 (2k_r^2-1)$, which is always positive since 
$c_{\rm ES}>0$ and $k_r>1$. 
Thus, these oscillatory tails always appear ahead of the embedded soliton. 
Behind the embedded soliton, there is the possibility that 
a flat shelf may develop as in the perturbed KdV equation 
\cite{karpman,kaupnewell} (see also \cite{PG2}). 
If a shelf develops, it moves to the region $x \ll -1$ 
at the velocity $-c_{\rm ES}$
in the moving coordinate system (\ref{movingsystem}). 

Thus, at large times $t \gg 1$, the boundary conditions
for the solution $u_1(\xi,t)$ are 
\begin{equation} \label{bc}
u_1(\xi, t) \longrightarrow \left\{ 
\begin{array}{ll}
R_{\rm rad}\sin(k_r\xi+\delta_{\rm rad}) H(c_{\rm gr} t-\xi), & \xi\gg 1,  \\
R_0 H(-\xi-c_{\rm ES} t), & \xi \ll -1,
\end{array}
\right. 
\end{equation}
where $R_{\rm rad}$ is the oscillatory-tail amplitude, 
$\delta_{\rm rad}$ is its phase, 
$R_0$ is the height of the trailing shelf, 
and $H(x)$ is the step function, i.e., 
$H=1$ when $x\ge 0$, and $H=0$ otherwise. 
Below, we determine the tail amplitude $R_{\rm rad}$ 
and the phase $\delta_{\rm rad}$. We also show 
that the shelf is not excited in the present 
situation, i.e., $R_0 = 0$. 

Our calculations of $R_0$, $R_{\rm rad}$ and $\delta_{\rm rad}$ 
are based on the observation that, 
as $t$ goes to infinity, the transient part of the solution $u_1(\xi, t)$
for Eq. (\ref{u1}) dies out, and $u_1(\xi, t)$ approaches
a steady state $u_{1s}(\xi)$ where
\begin{equation} \label{bcsteady}
u_{1s}(\xi) \longrightarrow \left\{
\begin{array}{ll}
R_{\rm rad}\sin(k_r\xi+\delta_{\rm rad}), & \xi\rightarrow \infty,  \\
R_0, & \xi \rightarrow -\infty. 
\end{array}
\right.
\end{equation}
This steady-state solution satisfies the same equation (\ref{u1}) except that
the time derivative in (\ref{u1}) is dropped, i.e., 
\begin{equation} \label{u1steady}
\left[{\cal L}u_{1s}\right]_{\xi}= U_{\rm ES}'(\xi). 
\end{equation}
Integration of this equation with respect to $\xi$ gives
\begin{equation} \label{u1steady2}
{\cal L}u_{1s}=U_{\rm ES}+\eta, 
\end{equation}
where $\eta$ is a constant. To determine $\eta$, we 
substitute the boundary condition (\ref{bcsteady}) of solution
$u_{1s}(\xi)$ at $\xi\gg 1$ into Eq. (\ref{u1steady2}) and find that
$\eta=0$. Then substitution of the boundary condition (\ref{bcsteady})
at $\xi\ll -1$ into Eq. (\ref{u1steady2}) readily shows that $R_0=0$. 
Thus, the flat shelf is not excited in the present situation. 

Since $R_0 = \eta = 0$, the inhomogeneous equation (\ref{u1steady2}) 
for $u_{1s}(\xi)$ becomes the same as Eq. (\ref{inhomo}). The general 
bounded solution for $u_{1s}(\xi)$ is 
\begin{equation} \label{u1solution}
u_{1s}(\xi)=\frac{\partial U}{\partial c}(\xi; c_{\rm ES}, \delta_0)
+\Gamma_1 \psi_1(\xi)+\Gamma_2\psi_2(\xi), 
\end{equation}
where $\delta_0$ is any fixed phase and $\Gamma_{1,2}$ are constants. 
The homogeneous solution $\psi_3(\xi)$ is excluded by 
a simple position normalization of the embedded soliton. 
The boundary condition of solution (\ref{u1solution}) at infinity 
can be obtained from Eqs. (\ref{tail}), (\ref{psi1}) and 
(\ref{psi2}) as 
\begin{equation}
u_{1s}(\xi) \longrightarrow \left\{ 
\begin{array}{ll}
\hspace{0.3cm} R(\delta_0)\sin(k_r \xi+\delta_0)+
\Gamma_1 \sin(k_r\xi+\delta_s)
         +\Gamma_2 \sin(k_r\xi+\delta_a),  & \xi\rightarrow \infty,  \\
-R(\delta_0)\sin(k_r\xi-\delta_0)-\Gamma_1 \sin(k_r\xi-\delta_s)
         +\Gamma_2 \sin(k_r\xi-\delta_a), & \xi \rightarrow -\infty. 
\end{array} \right. 
\end{equation}
This boundary condition should match condition 
(\ref{bcsteady}) with $R_0 = 0$. 
For convenience, we fix $\delta_0=\delta_a$. Then 
the matching condition gives 
the radiation amplitude $R_{\rm rad}$ and phase 
$\delta_{\rm rad}$ as 
\begin{equation} \label{tildeRdelta}
R_{\rm rad} = 2 R(\delta_a), \hspace{0.5cm} \delta_{\rm rad}=\delta_a, 
\end{equation}
and 
\begin{equation} \label{gamma12}
\Gamma_1=0,  \hspace{0.5cm} \Gamma_2=R(\delta_a). 
\end{equation}

Formulae (\ref{tildeRdelta}) are 
important results of this section. They show that the 
radiation phase $\delta_{\rm rad}$ is equal to the 
phase $\delta_a$ of the anti-symmetric homogeneous 
solution $\psi_2(\xi)$ [see Eq. (\ref{psi2})], while 
the radiation amplitude $R_{\rm rad} = 2 R(\delta_a)$. 
Since the minimal tail amplitude of symmetric nonlocal 
waves occurs at phase $\delta_{\rm min} = 
\left( \delta_s + \frac{\pi}{2} \right) \mbox{mod}(\pi)$ 
[see Eq. (\ref{deltamin})], and $\delta_a \ne \left( \delta_s + 
\frac{\pi}{2} \right) \mbox{mod}(\pi)$ in general
(see Sec. 4 for an example), we conclude that
the radiation amplitude $R_{\rm rad}$ generally is 
not minimal, i.e. $R_{\rm rad} \neq 2 R(\delta_{\rm min})$. 
In the numerical work for the KdV equation plus the
fifth-order derivative \cite{boydbook} (Sec. 16.6), it was mentioned 
without proof that radiation tail amplitude was minimal. 
That statement does not agree with our general analysis. 
But if the fifth-order KdV equation is integrable, then the relation 
$\delta_a = \delta_s + \frac{\pi}{2}$ holds \cite{yangstudy01}, 
i.e. the radiation amplitude is indeed minimal to the 
leading order of the perturbation theory 
for nearly integrable fifth-order KdV equations. 

When the first-order solutions (\ref{u1solution}) and (\ref{gamma12})
are substituted into the perturbation expansion (\ref{uexpand}), 
the solution can be re-written as 
\begin{equation}  \label{usolution2}
u(\xi, t)=\left\{U(\xi; c, \delta_a)
+(c-c_{\rm ES})R(\delta_a)\psi_2(\xi)+
{\rm O}[(c-c_{\rm ES})^2] \right\} H(c_{\rm gr} t-\xi), 
\hspace{0.5cm} t\gg 1,
\end{equation}
where $c(T)$ is given by Eq. (\ref{Vexpand}).
This solution describes the slow evolution 
of the perturbed embedded soliton in the fifth-order 
KdV equation (\ref{u}), while the fast radiation part 
produced by a general initial condition for $u(\xi,0)$ 
is neglected in the asymptotic limit $t \gg 1$.
Solution (\ref{usolution2}) up to order 
${\rm O}(c-c_{\rm ES})$ consists of a symmetric 
nonlocal wave $U(\xi;c,\delta_{\rm a})$ and an 
anti-symmetric term $\psi_2(\xi)$. This 
anti-symmetric term is generated 
in the initial-value evolution problem due to 
the (radiation) boundary condition (\ref{bc})
with $R_0 = 0$, $R_{\rm rad} = 2 R(\delta_a)$, and
$\delta_{\rm rad} = \delta_a$. 
Since function $\psi_2(\xi)$ is anti-symmetric, the 
radiation amplitude is canceled behind the embedded 
soliton and is doubled ahead of the soliton.
It is also noted that $\psi_2(0)=0$, thus the amplitude 
of solution (\ref{usolution2}) at soliton center
$\xi=0$ is then the same as that of
the symmetric nonlocal wave $U(\xi; c, \delta_{\rm a})$. 
This fact will be used in Sec. 4 in our comparison between 
the analytical and numerical results on 
the amplitudes of perturbed embedded solitons. 

When the radiation amplitude $R_{\rm rad}$ and phase 
$\delta_{\rm rad}$ are found, we are
ready to derive the dynamical equation for the velocity 
$c(T)$ of a perturbed embedded soliton. This equation can be derived
in several different ways (see \cite{yangstudy01}). 
The simplest way is to use the local or global momentum conservation law 
when the system (\ref{u}) is Hamiltonian. 
The derivation using the global momentum conservation law 
(\ref{momentum}) is presented below. 
The derivation using the local momentum conservation
is contained in Appendix B. 

To derive the velocity equation, we  substitute 
the perturbation expansion (\ref{uexpand})
into the momentum integral (\ref{momentum}). 
When terms up to order $\epsilon^2$ are retained, we get 
\begin{equation} \label{momentum2}
\frac{d}{dt}\int^{\infty}_{-\infty} \left\{U_{\rm ES}^2+2\epsilon 
U_{\rm ES}(c_1 u_1+\epsilon u_2)+\epsilon^2 c_1^2 u_1^2\right\}d\xi=0.
\end{equation}
Keep in mind that solutions $u_1(\xi,t)$ and $u_2(\xi,t)$ 
at the center region $\xi\sim O(1)$ become stationary as 
$t \gg 1$. As a result, the term involving $u_2$ in 
Eq. (\ref{momentum2}) can be dropped because the integral of the
product $U_{\rm ES}(\xi) u_{2}(\xi,t)$ becomes constant at large times. 
The stationary solution $u_{1s}(\xi)$ is given by 
Eqs. (\ref{u1solution}) and (\ref{gamma12}). Thus,
\begin{equation} \label{integral}
E \equiv \int^{\infty}_{-\infty} U_{\rm ES} u_{1s} d\xi=
\int^{\infty}_{-\infty} U_{\rm ES}(\xi) 
\frac{\partial U}{\partial c}(\xi; c_{\rm ES}, \delta_a) d\xi. 
\end{equation}
Lastly, the solution for $u_1(\xi,t)$ develops an oscillatory 
tail ahead of the embedded soliton. This tail has amplitude 
$R_{\rm rad}$ given by Eq. (\ref{tildeRdelta}), and 
it moves at its group velocity $c_{\rm gr}$. 
The average energy ($u_1^2$) of the sinusoidal tail 
is $\frac{1}{2}R_{\rm rad}^2$, i.e., $2R^2(\delta_a)$. Thus, 
\begin{equation} \label{radiationderiv}
\Gamma \equiv \frac{d}{dt}\int^{\infty}_{-\infty} u_1^2 d\xi=
2R^2(\delta_a) c_{gr}=
4k_r^2(2k_r^2-1)R^2(\delta_a). 
\end{equation}
When relations (\ref{integral}) and (\ref{radiationderiv}) 
are substituted into the momentum equation (\ref{momentum2}) 
and $\epsilon c_1$ replaced by $c-c_{\rm ES}$ 
[see Eq. (\ref{Vexpand})], we finally obtain the dynamical 
equation for the perturbed embedded-soliton's velocity $c$ as
\begin{equation} \label{c}
\frac{dc}{dt}=-\beta (c-c_{\rm ES})^2, 
\end{equation}
where the coefficient $\beta = \Gamma /2E$. 
The solution of Eq. (\ref{c}) is
\begin{equation}\label{solution}
c(t)=c_{\rm ES}+\frac{c_0-c_{\rm ES}}{1+\beta (c_0-c_{\rm ES})t}, 
\end{equation}
where $c_0$ is the initial condition for velocity $c(t)$. 
The formula for $c_0$ can be obtained from Eqs. (\ref{Vexpand})
and (\ref{projection}) as 
\begin{equation} \label{c0}
c_0=c_{\rm ES}+\frac{\int_{-\infty}^\infty U_{\rm ES}(\xi) 
\left[ u(\xi, 0) - U_{\rm ES}(\xi) \right] d\xi}{ 
\int_{-\infty}^{\infty} U_{\rm ES}(\xi) 
\frac{\partial U}{\partial c}(\xi;c_{\rm ES},\delta_{\rm a})d\xi}. 
\end{equation}
Once the initial perturbed embedded soliton $u(\xi,0)$ is specified, 
then $c_0$ is fixed as above. 

The asymptotic equation (\ref{c}) is the key result of this paper. 
When $\beta>0$, this equation shows that its fixed point 
$c=c_{\rm ES}$ is semi-stable: any perturbation with 
$c_0>c_{\rm ES}$ is stable, and any perturbation with 
$c_0<c_{\rm ES}$ is unstable. When translated into the 
original partial differential equation (\ref{u}), 
it means that the embedded soliton is semi-stable. 
Depending on the type of initial perturbations, the embedded soliton
can persist, or be destroyed. 

Finally, when $E$ vanishes, the zero embedded eigenvalue 
has multiplicity higher than two, which results in linearized 
(algebraic) instability of the embedded soliton (see, e.g., 
\cite{PG2}). We have excluded such linearized 
instability in our assumptions above.

\section{Comparison with direct numerical simulations}

In this section, we directly simulate the original partial 
differential equation (\ref{u}), and compare the results 
with our analytical theory above. 
The system parameters we choose are 
\begin{equation} \label{parameters}
\alpha_0=5, \hspace{0.2cm} 
\alpha_1=5, \hspace{0.2cm} 
\alpha_2=2.5, \hspace{0.2cm} 
\alpha_3=0 
\end{equation}
in Eq. (\ref{N}). Note that these parameter values 
are equivalent to $\alpha_0=\alpha_1=1, \alpha_2=0.5$ and $\alpha_3=0$
after variable $u$ and time $t$ are rescaled. 
At these parameter values, Eq. (\ref{u}) is Hamiltonian. 
The fifth-order long-wave model equation studied by
Champlneys and Groves \cite{groves} corresponds to our equation
(\ref{u}) with $\alpha_0 = 1$ and $\alpha_3 = 0$.
We have also tested other parameter values with $\alpha_1 =2 \alpha_2$
and found similar results. 
For instance, in the third-order Hamiltonian long-wave approximation
to the water-wave problem as derived by Craig and Groves \cite{craig}, 
the parameter values (after variable rescaling) are
$\alpha_0=1, \alpha_1=-\frac{5}{3}, \alpha_2=-\frac{5}{6}$ 
and $\alpha_3=0$. Comparison between
our theory and numerics for this set of parameters 
is qualitatively the same as that for the parameters (\ref{parameters}). 

\begin{figure}[p]
\begin{center}
\parbox{8cm}{\postscript{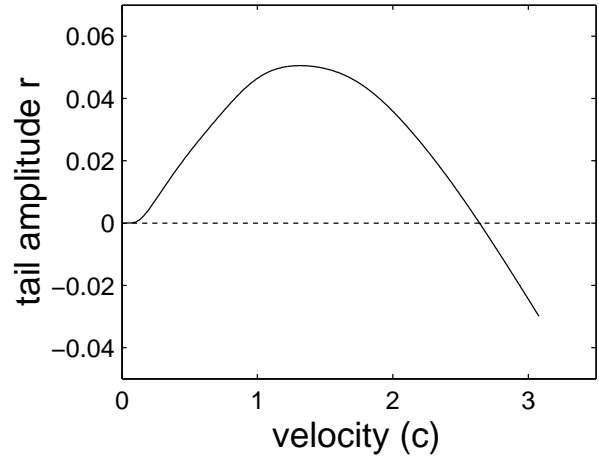}{1.0}}

\caption{Tail amplitude $r(c,\delta_a)$ of symmetric nonlocal waves
as a function of velocity $c$. The nonlocal waves satisfy 
Eq. (\ref{ode}) with system parameters (\ref{parameters}). 
The radiation phase $\delta = \delta_a$ is the phase of 
the anti-symmetric linear mode $\psi_2$ given by Eq. (\ref{psi2}): 
$\delta_a \approx 0.5737$. \label{rcurve} }
\end{center}
\end{figure}

With the parameters (\ref{parameters}), the fifth-order KdV equation
(\ref{u}) has an embedded soliton
\begin{equation} \label{ES}
U_{\rm ES}(x, t)=0.9 \:\mbox{sech}^2\left[\sqrt{0.3}
\hspace{0.05cm} (x-c_{\rm ES}t)\right]
\end{equation}
at the exact wave speed $c_{\rm ES}=2.64$ (see \cite{champneysphysD01}). 
The approximate phase values $\delta_s$ and $\delta_a$ in the linear modes
$\psi_{1,2}$ of the linearization operator ${\cal L}$
are found numerically (by the shooting method) as 
\begin{equation}
\delta_s=2.1815, \hspace{0.5cm} 
\delta_a=0.5737. 
\end{equation}
Note that the difference between these two phases here is not equal 
to $\pi/2$, thus the radiation tail amplitude in perturbed embedded 
solitons is not minimal. 
However, this phase difference differs from $\pi/2$ only by 0.037. 
Thus the radiation tail amplitude is rather close to its minimal value. 
At phase $\delta=\delta_a$ (which is the radiation phase), 
we have numerically obtained the tail amplitude curve $r(c, \delta_a)$
of symmetric nonlocal waves from Eq. (\ref{ode}) at various velocity 
$c$ values, again by the shooting method. The results are shown 
in Fig. \ref{rcurve}. As expected, the tail amplitude $r$ is 
non-zero for $c>0$ except when $c=c_{\rm ES}$. The slope 
$R(\delta_a)$ at embedded-soliton velocity $c_{\rm ES}$
is found to be $R(\delta_a)=-0.0652$. The $k_r$ value can be quickly
obtained from Eq. (\ref{k}), and the integral in Eq. (\ref{integral}) can
be readily determined numerically. From all these values, we finally 
found that $\beta=0.0868$. With this $\beta$ value, 
our analytical formula for the velocity of perturbed
embedded solitons is then given by Eq. (\ref{solution}). 

In order to verify our analytical theory, we have numerically 
simulated the original wave equation (\ref{u}) with system
parameters (\ref{parameters}) and initial condition
\begin{equation} \label{ic}
u(x, 0)=h\: U_{\rm ES}(x, 0), 
\end{equation}
where $h$ is a constant coefficient. Note that 
$h=1$ gives the exact embedded soliton, and $h\ne 1$ gives a 
perturbed embedded soliton. 
Our numerical scheme is the integrating factor method
as described in \cite{milewsky}. 
The $x$ interval is taken as 400 units long, and 1024 grid points
are used. The time stepsize is $10^{-4}$. 
To prevent radiation from re-entering the simulation
region through periodic boundary conditions, we have used a damping
condition near the boundaries. 
In our simulation, we have also adopted a frame moving at
the embedded-soliton's velocity $c_{\rm ES}$ (but the results will still
be presented in the original frame). 
Our numerical scheme has been tested with the exact embedded soliton 
(\ref{ES}) as the initial condition. It has also been tested 
on a related system --- the integrable fifth-order KdV hierarchy equation. 
Furthermore, we have tried different grid points and time stepsizes. 
These tests show that the numerical error in our scheme is on the order
of $10^{-6}$. 

\begin{figure}[p]
\begin{center}
\parbox{14cm}{\postscript{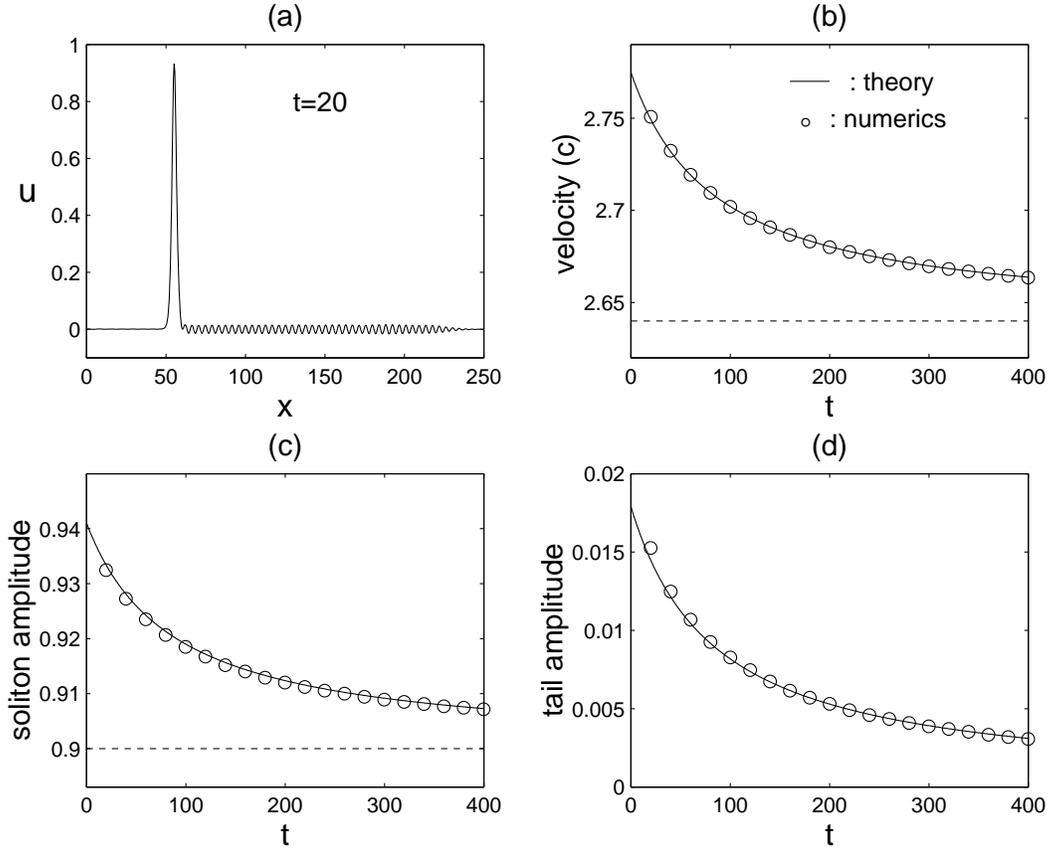}{1.0}}

\caption{Evolution of the embedded soliton under 
the momentum-enhancing perturbation (\ref{ic}) with $h=1.05$:
(a) numerical solution profile at $t=20$; 
(b) velocity $c$ of the perturbed embedded-soliton versus time $t$, 
the dashed line is the embedded-soliton velocity $c_{\rm ES} = 2.64$; 
(c) amplitude of the perturbed embedded-soliton versus time $t$, 
the dashed line is the embedded-soliton amplitude 0.9;
(d) radiation tail amplitude versus time $t$. 
The solid lines in (b,c,d) are theoretical predictions, 
and circles are numerical values.
\label{h105}   }
\end{center}
\end{figure}

\begin{figure}[p]
\begin{center}
\parbox{14cm}{\postscript{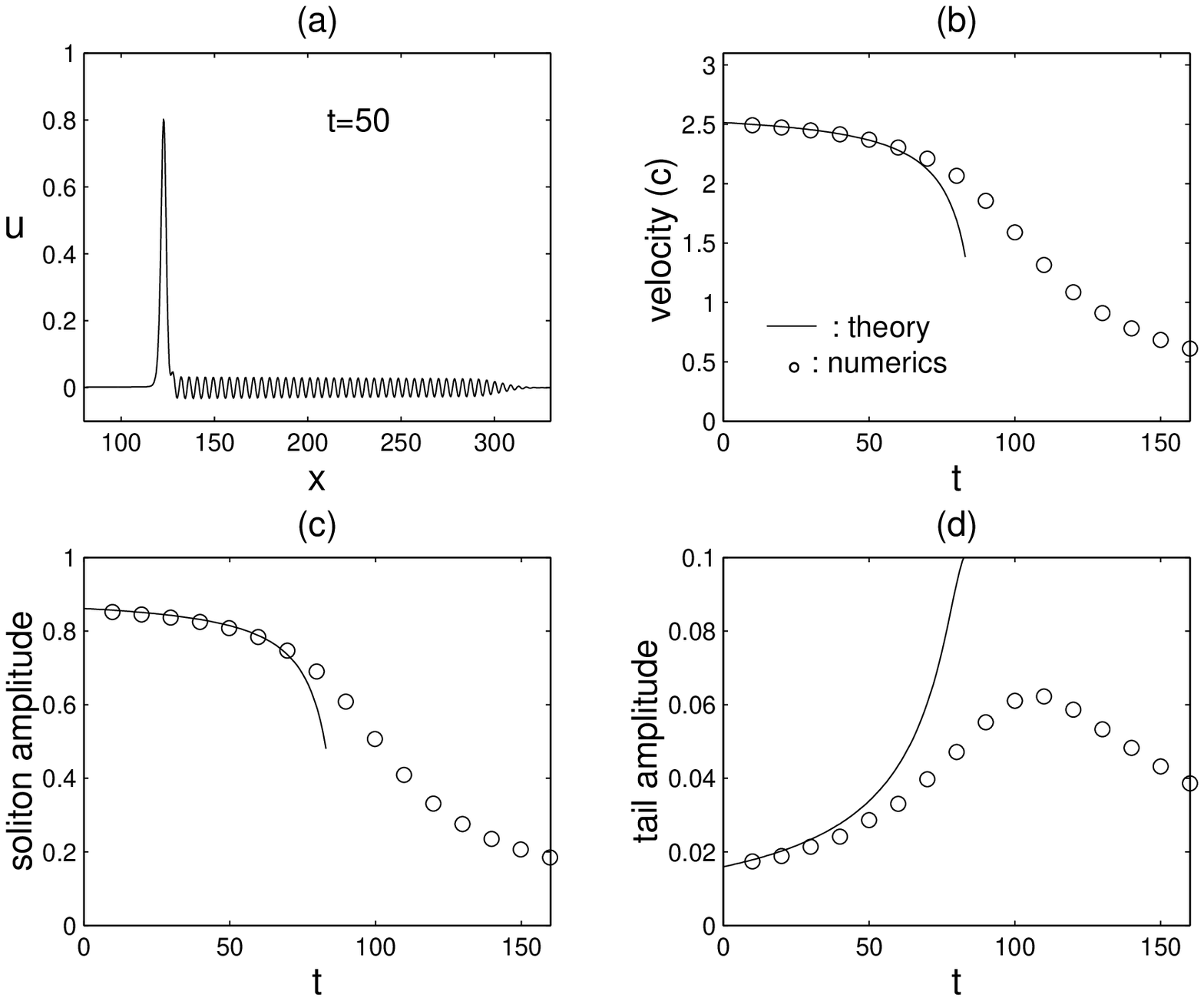}{1.0}}

\caption{Evolution of the embedded soliton under 
the momentum-reducing perturbation (\ref{ic}) with $h=0.95$:
(a) numerical solution profile at $t=50$; 
(b) velocity $c$ of the perturbed embedded-soliton versus time $t$; 
(c) amplitude of the perturbed embedded-soliton versus time $t$;
(d) radiation tail amplitude versus time $t$. 
The solid lines in (b,c,d) are theoretical predictions, 
and circles are numerical values.
\label{h095}}
\end{center}
\end{figure}

We have run our numerical scheme on two typical initial conditions 
(\ref{ic}) with $h=1.05$ and $0.95$. 
The results are presented in Figs. \ref{h105} and \ref{h095} respectively. 
In the former case, the perturbed state has momentum $P$ higher than 
the embedded soliton's [see Eq. (\ref{momentum})]. Because of this, the 
perturbed state initially
moves a little faster than the unperturbed embedded soliton 
[see Fig. \ref{h105}(b)]. But its speed as well as amplitude 
slowly decrease due to 
continuous wave radiation which moves ahead of the main pulse
[see Fig. \ref{h105}(b,c)]. 
This tail radiation at $t=20$ is shown in 
Fig.  \ref{h105}(a) (note that the tail decay near the right end
of the $x$-interval is due to our damping boundary condition. The actual
tail length is much longer). 
But the tail amplitude decreases also in the process [see Fig. \ref{h105}(d)]. 
Thus energy radiation is decreasing. Eventually the perturbed state 
asymptotically approaches the unperturbed embedded soliton, which is 
clearly seen in Fig. \ref{h105}(b,c). 

When $h=0.95$, the perturbed state has momentum $P$ lower than 
the embedded soliton's. In this case, due to 
continuous wave radiation which intensifies over time
[see Fig. \ref{h095}(a,d)], the speed and amplitude of the perturbed
embedded soliton both decrease well below their corresponding values
of the unperturbed embedded soliton [see Fig. \ref{h095}(b,c)]. 
When the amplitude of 
the main pulse has dropped significantly, it can no longer sustain
high radiation tails. Thus tail amplitudes start to decrease
[see Fig. \ref{h095}(d)]. Eventually, the embedded soliton 
is destroyed by perturbations. 

The above numerical simulation results agree both qualitatively and
quantitatively with our analytical theory. 
Qualitatively, when $h=1.05$, as the initial velocity is above 
$c_{\rm ES}$, formula (\ref{solution}) predicts that the pulse
velocity will asymptotically approach $c_{\rm ES}$; when 
$h=0.95$, the velocity will decay far below $c_{\rm ES}$. This 
semi-stability behavior is accurately reflected in the 
numerical results. 
Quantitatively, we have also compared the pulse velocity, amplitude 
and tail amplitude of analytical predictions to those of numerical results. 
The analytical prediction for pulse velocity is given by 
formula (\ref{solution}). The initial condition $c_0$ is calculated from
formula (\ref{c0}). We found that when $h=1.05$, 
$c_0\approx 2.775$, and when $h=0.95$, $c_0\approx 2.515$. 
The analytical prediction for the pulse amplitude is 
the center amplitude of symmetric nonlocal waves at analytically
predicted velocity $c$ [see Eq. (\ref{usolution2})]. 
The analytical prediction for tail amplitude is 
$(c - c_{\rm ES}) R_{\rm rad}$, where $R_{\rm rad}$ is 
given by Eq. (\ref{tildeRdelta}). 
These analytical predictions have been plotted in 
Figs. \ref{h105}(b,c,d) and \ref{h095}(b,c,d) as well for comparison. 
In the case $h=1.05$, the quantitative agreement between theory 
and numerics is excellent at all times. In the other case $h=0.95$, the 
quantitative agreement is good at the beginning, and gets worse 
at larger times. The good agreement in the former case is because 
the main pulse remains close to the embedded soliton at all times, 
thus the perturbation theory works well. 
In the latter case, the main pulse deviates significantly from the 
embedded soliton at large times. When that happens, the 
perturbation theory breaks down.

\section{Conclusion}

In this article, we have studied the evolution of perturbed 
embedded solitons in a general Hamiltonian fifth-order KdV
equation (\ref{u}). We have shown that when an embedded soliton
is perturbed, it sheds continuous-wave radiation in front of
the embedded soliton. The amplitude of this continuous wave
is not minimal in general. 
Behind the embedded soliton, no flat shelf is created. 
We have further derived 
the velocity equation of a perturbed embedded soliton. As a result, 
the semi-stability property of embedded solitons is analytically 
proved. In addition, we have obtained the conditions under which
a perturbed embedded soliton will decay or persist. 
We have also simulated the fifth-order KdV equation 
numerically. The numerical results agree well with the analysis 
both qualitatively and quantitatively. 

The analysis and the final dynamical equation (\ref{c}) 
are similar to those found in \cite{peliyang} for generalized 
SHG models. Thus, in spite of differences in the 
spectral properties of linearization operators in these two models, 
the nonlinear resonance between the embedded soliton 
and the continuous-wave radiation has common features 
under assumptions listed in Section 3. Obviously, the 
same method can be applied to any other embedded-soliton-bearing 
Hamiltonian system. In fact, the system does not even have 
to be Hamiltonian. A non-trivial conservation law 
such as power or momentum would be sufficient
to guarantee the semi-stability property of embedded solitons
(see \cite{peliyang}). The open problems going beyond the 
present study include generation and collisions of several 
embedded solitons, as well as further engineering applications 
of embedded solitons in applied science. 

\section*{\hspace{0.1cm} Acknowledgments}

The work of Y.T and J.Y. was supported in part by 
the Air Force Office of Scientific Research under 
contract F49620-99-1-0174, and by the National Science 
Foundation under grant DMS-9971712. 
The work of D.P. was supported by NSERC grant 5-36694.

\section*{\hspace{0.1cm} Appendix A}
\def\theequation{A.\arabic{equation}}

In this appendix, we present a more rigorous approach for solving
the inhomogeneous equation (\ref{u1}) for the first-order 
solution $u_1(\xi,t)$ by using a spectral decomposition method 
for the linearized problem, 
\begin{equation} \label{linearproblem}
({\cal L} \phi(\xi;k))_{\xi} = i \Omega(k) \phi(\xi;k),
\end{equation} 
where $\Omega(k) = k ( k^4 - k^2 - c_{\rm ES} )$,
and $\phi(\xi;k)$ are continuous-wave eigenfunctions normalized 
according to the boundary condition:
\begin{equation} \label{bc1}
\phi(\xi;k) \longrightarrow e^{i k \xi}, \;\;\;{\rm as}\;\;\; 
\xi \ll -1.
\end{equation}
The potential terms with $U_{\rm ES}(\xi)$
decay exponentially at large $|\xi|$ in the operator
${\cal L}$ given by Eq. (\ref{L}). 
The eigenfunctions $\phi(\xi,k)$ may have up to three Fourier oscillatory 
terms in the limit $\xi \to + \infty$, which match with the roots 
of the equation: $\Omega(k) = \Omega$. 
We will compute asymptotically the Fourier-type integrals
[see Eq. (\ref{analysis}) below] 
at the resonant values $k = \pm k_r$, where $k_r \equiv k(c_{\rm ES})$ 
and $k(c)$ is given by Eq. (\ref{k}). At the resonant values, 
the eigenvalue parameter $\Omega$ is zero, i.e. $\Omega(\pm k_r) = 0$. 
It could be found from Eqs. (\ref{psi1}) and (\ref{psi2}) that the 
boundary condition for $\phi(\xi,\pm k_r)$ in the limit $\xi \to +\infty$ 
is:
\begin{equation} \label{bc2}
\phi(\xi;\pm k_r) \longrightarrow a_{\pm} e^{\pm i k_r \xi} 
+ b_{\pm} e^{\mp i k_r \xi} + c_{\pm}, \;\;\;{\rm as}\;\;\; 
\xi \gg 1,
\end{equation}
where 
$$
a_{\pm} = \frac{e^{\pm i (\delta_s + \delta_a - \pi/2)}}{
\sin(\delta_s-\delta_a)}, 
\;\;\;\;
b_{\pm} = \frac{\pm i \cos(\delta_s - \delta_a)}{\sin(\delta_s-\delta_a)}, 
\;\;\;\; c_{\pm} = 0.
$$
Under assumptions described in Section 3, the solution 
for $u_1(\xi,t)$ can be decomposed through eigenfunctions 
of the linearized problem (\ref{linearproblem}):
\begin{equation} \label{decomposition}
u_1(\xi,t) = \frac{\partial U}{\partial c} (\xi;c_{\rm ES},\delta) 
+ \int_{-\infty}^{\infty} w(k,t) \phi(\xi,k) dk + \alpha U'_{\rm ES}(\xi),
\end{equation}
where $\alpha$ is constant. The first term in (\ref{decomposition}) 
solves the inhomogeneous part of Eq. (\ref{u1}). Since 
the double eigenvalue $\Omega = 0$ 
is embedded into the integral at the resonant points $k = \pm k_r$, 
the inhomogeneous term is not independent and can be 
decomposed through the same eigenfunctions:
\begin{equation} \label{decomposition1}
\frac{\partial U}{\partial c}(\xi;c_{\rm ES},\delta) = 
\int_{-\infty}^{\infty} \frac{F(k) \phi(\xi,k)}{k^2 - k_r^2} dk,
\end{equation}
where $F(\pm k_r) \ne 0$. The singular (pole) part 
in (\ref{decomposition1}) describes the non-localized 
oscillatory tail (\ref{asym2}) as $|\xi| \gg 1$.
With Eq. (\ref{decomposition1}) substituted into Eq. (\ref{decomposition}), 
at time $t=0$, Eq. (\ref{decomposition}) becomes
\begin{equation}
u_1(\xi,0) = \int_{-\infty}^{\infty} \tilde{w}(k,0) \phi(\xi,k) dk + \alpha U'_{\rm ES}(\xi),
\end{equation}
where
\begin{equation}
\tilde{w}(k,0)=\frac{F(k)}{k^2-k_r^2}+w(k,0).
\end{equation}
When the initial condition $u_1(\xi,0)$ is localized, then the
spectral component $\tilde{w}(k,0)$ is free of pole singularities. 

The complex amplitude $w(k,t)$ in Eq. (\ref{decomposition}) 
satisfies the trivial evolution equation:
\begin{equation}
\label{trivialevolution}
\frac{\partial w}{\partial t} + i \Omega(k) w=0. 
\end{equation}
Solving the initial-value inhomogeneous problem (\ref{u1}) 
with the spectral decompositions (\ref{decomposition}) and 
(\ref{decomposition1}), we find the following integral 
representation for $u_1(\xi,t)$:
\begin{equation} \label{w}
u_1(\xi,t) = \int_{-\infty}^{\infty} \frac{F(k) 
\phi(\xi,k)}{k^2 - k_r^2} \left( 1 - e^{-i \Omega(k) t} \right) dk 
+ \int_{-\infty}^{\infty} \tilde{w}(k,0) e^{-i \Omega(k) t} \phi(\xi,k) dk 
+ \alpha U'_{\rm ES}(\xi). 
\end{equation}
The second term in Eq. (\ref{w}) represents the non-singular 
part produced by the initial condition $u_1(\xi,0)$.
The first term in Eq. (\ref{w}) represents the singular 
(pole) part produced by slow evolution of the embedded soliton. 
The singular integral term describes the nonlinear resonance 
between the embedded soliton and the continuous-wave radiation. 

The singular (pole) term in Eq. (\ref{w}) 
occurs for $k = \pm k_r$, i.e. when $\Omega(\pm k_r) = 0$. 
This term represents the oscillatory-tail radiation that 
diverges from the embedded soliton with the group velocity 
$c_{\rm gr} = \Omega'(k_r) = 2k_r^2(2k_r^2-1)$. Since 
$c_{\rm ES} > 0$ and $k_r^2 > 1$, then $c_{\rm gr} > 0$, 
i.e. the oscillatory-tail radiation occurs ahead but 
not behind of the embedded soliton. We prove this conjecture 
by using the pole decomposition technique \cite{peliyang}. 
The singular contribution from the integral (\ref{w}) 
can be evaluated in the asymptotic region $\xi \ll -1$, 
$t \gg 1$ such that $|\xi|/t$ is constant:
\begin{eqnarray}
\nonumber
\lim_{t \to + \infty} 
\int_{-\infty}^{\infty} \frac{F(k) dk}{(k^2 - k_r^2)} 
e^{i k \xi} \left( 1  - e^{-ik(k^4 - k^2 - c_{\rm ES})t} \right) \\
\label{analysis}
= \frac{\pi i}{2 k_r} \left( 
F(k_r) e^{i k_r \xi} - \bar{F}(k_r) e^{-i k_r \xi} \right) 
\left[ {\rm sign}(\xi/t) - {\rm sign}(\xi/t - c_{\rm gr}) \right] = 0. 
\end{eqnarray}
Here we have used the boundary condition (\ref{bc1}) and the symmetry 
relation $\bar{F}(k_r) = F(-k_r)$. Similar but lengthy computations 
of the integral (\ref{w}) with the boundary condition (\ref{bc2}) 
in the region $\xi \gg 1$ prove that the boundary conditions 
for the solution $u_1(\xi,t)$ of Eq. (\ref{u1}) in the limit 
$t \gg +\infty$ and $\xi/t$ constant is 
\begin{equation} \label{bcdmitry}
u_1(\xi, t) \longrightarrow \left\{ 
\begin{array}{ll}
R_{\rm rad} \sin(k_r \xi + \delta_{\rm rad}) 
H( c_{\rm gr} t-\xi), & \xi\gg 1,  \\
0, & \xi \ll -1,
\end{array}
\right. 
\end{equation}
where $R_{\rm rad}$ is the radiation amplitude, 
\begin{equation}
\label{alterR}
R_{\rm rad} = \frac{2 \pi |F(k_r)|}{k_r \sin(\delta_a -\delta_s)}, 
\end{equation}
$\delta_{\rm rad}$ is the radiation phase, 
\begin{equation}
\label{alterD}
\delta_{\rm rad} = \arg(F(k_r)) + \delta_s + \delta_a - \frac{\pi}{2},
\end{equation}
and $H(x)$ is the step function, i.e., $H=1$ when $x\ge 0$, 
and $H=0$ otherwise. 

We show that these results are consistent with Eqs. 
(\ref{bc}) and (\ref{tildeRdelta}). Indeed, computing 
the singular contribution from the integral (\ref{decomposition1}) 
in the region $\xi \ll -1$, we find by similar technique that:
\begin{equation}
\label{computationIntegral}
\int_{-\infty}^{\infty} \frac{F(k) dk}{(k^2 - k_r^2)} 
e^{i k \xi} = - \frac{\pi i}{2 k_r} \left( 
F(k_r) e^{i k_r \xi} - \bar{F}(k_r) e^{-i k_r \xi} \right). 
\end{equation}
Matching this boundary condition with Eq. (\ref{asym2}) 
in the region $\xi \ll -1$, we find:
\begin{equation}
\label{matchingrelations}
|F(k_r)| = \frac{k_r}{\pi} R(\delta), \;\;\;\;
\arg(F(k_r)) = \pi - \delta.
\end{equation}
Let us specify the inhomogeneous solution 
$\partial U(\xi;c_{\rm ES},\delta) / \partial c$ 
at the minimal tail phase $\delta = \delta_s + \pi/2$, 
then the homogeneous eigenfunction $\psi_1(\xi)$ is 
excluded from (\ref{linearsolution}) and (\ref{decomposition1}), 
since $\gamma_1 = 0$, see below Eq. (\ref{RRRRR}). 
Setting the value $\delta = \delta_s + \pi/2$ in 
Eq. (\ref{matchingrelations}), we finally find from 
Eqs. (\ref{alterR}) and (\ref{alterD}) that 
$$
R_{\rm rad} = 2 \frac{R(\delta_s + 
\frac{\pi}{2})}{\sin(\delta_a -\delta_s)} = 2 R(\delta_a), \;\;\;\;
\delta_{\rm rad} = \delta_{\rm a},
$$
where we have used the relation (\ref{RRRRR}). 

\section*{\hspace{0.1cm} Appendix B}
\def\theequation{B.\arabic{equation}}
\setcounter{equation}{0}

In this appendix, we use a local momentum conservation law to derive
the dynamical equation for velocity $c$. 
The local momentum conservation law has the form,
\begin{equation} \label{local}
\left[ \frac{1}{2} u^2 \right]_t + 
\left[ u u_{xx} - \frac{1}{2} u_x^2 + u u_{xxxx} - u_x u_{xxx} 
+ \frac{1}{2} u_{xx}^2 + \frac{2}{3} \alpha_0 u^3 
+ \alpha_1 u^2 u_{xx} + \frac{3}{4} \alpha_3 u^4 \right]_x 
= 0.
\end{equation}
Integrating Eq. (\ref{local}) over $|\xi| \sim {\rm O}(1)$ 
and substituting the perturbation expansion (\ref{uexpand})
up to order ${\rm O}(\epsilon^2)$, we derive the following 
equation, 
\begin{equation} \label{momentum3}
\left( \int^{\infty}_{-\infty} U_{\rm ES} u_1 d \xi \right) 
\frac{d c_1}{d T} + c_1^2 \left[ 
- \frac{1}{2} c_{\rm ES} u_1^2 + u_1 u_{1 \xi \xi} - \frac{1}{2} 
u_{1 \xi}^2 + u_1 u_{1 \xi \xi \xi \xi} - u_{1 \xi} 
u_{1 \xi \xi \xi} + \frac{1}{2} u_{1 \xi \xi}^2 \right]_{\xi 
\to - \infty}^{\xi \to + \infty} = 0.
\end{equation}
The stationary solution $u_{1s}(\xi)$ 
is given by Eqs. (\ref{u1solution}) and (\ref{gamma12}), 
and the oscillatory tail in front of the embedded soliton is
given by Eqs. (\ref{bc}) and (\ref{tildeRdelta}).
When those formulas are utilized, we get 
\begin{equation} \label{integral1}
\int^{\infty}_{-\infty} U_{\rm ES}u_{1} d\xi=E, 
\end{equation}
and 
\begin{equation} \label{radiationderiv2}
\left[ - \frac{1}{2} c_{\rm ES} u_1^2 + u_1 u_{1 \xi \xi} - \frac{1}{2} 
u_{1 \xi}^2 + u_1 u_{1 \xi \xi \xi \xi} - u_{1 \xi} 
u_{1 \xi \xi \xi} + \frac{1}{2} u_{1 \xi \xi}^2 \right]_{\xi 
\to - \infty}^{\xi \to + \infty} = 2 k_r^2 ( 2 k_r^2 - 1 ) R^2(\delta_a)
=\frac{1}{2} \Gamma, 
\end{equation}
where $E$ and $\Gamma$ are defined in Eqs. (\ref{integral}) and
(\ref{radiationderiv}). 
When relations (\ref{integral1}) and (\ref{radiationderiv2}) 
are substituted into Eq. (\ref{momentum3}) and $\epsilon c_1(T)$ 
is replaced by $c(t) - c_{\rm ES}$ (\ref{Vexpand}), 
the dynamical equation (\ref{c}) is reproduced.

\begin{thebibliography}{200}

\bibitem{olver} S. Kichenassamy and P.J. Olver,  
Existence and nonexistence of solitary wave solutions 
to higher-order model evolution equations,
SIAM J. Math. Anal. {\bf 23} (1992), 1141. 

\bibitem{akylas} D.C. Calvo and T.R. Akylas, 
 On the formation of bound states by interacting 
nonlocal solitary waves, Physica D 101 (1997), 270. 

\bibitem{groves} A.R. Champneys and M.D. Groves, 
 A global investigation of solitary wave
solutions to a two-parameter model for water waves, 
J. Fluid Mech. {\bf 342} (1997), 199. 

\bibitem{buryak} A.V. Buryak,  Stationary soliton bound 
states existing in resonance with linear waves, 
Phys. Rev. E 52 (1995), 1156. 

\bibitem{fujioka} J. Fujioka and A. Espinosa,  Soliton-like
solution of an extended NLS equation existing in resonance with
linear dispersive waves, J. Phys. Soc. Japan 66 (1997), 2601. 

\bibitem{grimshaw} R. Grimshaw and P. Cook,  Solitary waves 
with oscillatory tails, in {\em Proceedings of the Second 
International Conference on Hydrodynamics}, editors: 
A.T. Chwang, J.H.W. Lee, and D.Y.C. Leung, Hong Kong, 1996. 

\bibitem{yangprl99} J. Yang, B.A. Malomed, and D.J. Kaup, 
 Embedded solitons in second-harmonic-generating systems, 
Phys. Rev. Lett. {\bf 83} (1999), 1958. 

\bibitem{thirring} A.R. Champneys, and B.A. Malomed, and M.J. Friedman, 
 Thirring solitons in the presence of dispersion, 
Phys. Rev. Lett. {\bf 80} (1998), 4168. 

\bibitem{movingES} A.R, Champneys and B.A. Malomed,  
 Moving embedded solitons, J. Phys. A 32 (1999), L547. 

\bibitem{threewave} A.R. Champneys and B.A. Malomed, 
 Embedded solitons in a three-wave system, 
Phys. Rev. E 61 (1999), 886. 

\bibitem{champneysphysD01} A.R. Champneys, B.A. Malomed,  
J. Yang, and D.J. Kaup,  Embedded solitons: solitary waves 
in resonance with the linear spectrum, Physica D 152 (2001), 340. 

\bibitem{yangathens01} J. Yang, B.A. Malomed, D.J. Kaup, 
and A.R. Champneys,  Embedded solitons: a new type of 
solitary waves, Mathematics and Computers in Simulation 
{\bf 56} (2001), 585. 

\bibitem{yangstudy01} J. Yang,  Dynamics of embedded 
solitons in the extended KdV equations, 
Stud. Appl. Math. 106 (2001), 337. 

\bibitem{peliyang} D.E. Pelinovsky and J. Yang, 
 A normal form for nonlinear resonance of embedded solitons,
Proc. Roy. Soc. Lond. A (2002), to be published. 

\bibitem{DK} F. Dias and E.A. Kuznetsov,  Nonlinear 
stability of solitons in the fifth-order Korteweg--de Vries 
equation, Phys. Lett. A {\bf 263} (1999), 98. 

\bibitem{Levandovsky} S.P. Levandovsky,  A stability 
analysis for firth-order water-wave models, 
Physica D {\bf 125} (1999), 222. 

\bibitem{bridges} T.J. Bridges and G. Derks,  The 
symplectic Evans matrix, and the instability of solitary 
waves and fronts, Arch. Rational Mech. Anal. {\bf 156} (2001), 
1. 

\bibitem{PG2} D.E. Pelinovsky and R.H.J. Grimshaw,  An asymptotic
approach to solitary wave instability and critical collapse in 
long-wave KdV-type evolution equations.
Physica D {\bf 98} (1996), 139. 

\bibitem{champneys} A.R. Champneys, 
 Codimension-one persistence beyond all orders of 
homoclinic orbits to singular saddle centres in 
reversible systems, Nonlinearity {\bf 14} (2001), 87-112.

\bibitem{benilovmalomed} R. Grimshaw, B. Malomed, and 
E.S. Benilov,  Solitary waves with damped oscillatory 
tails: an analysis of the fifth-order Korteweg-de 
Vries equation, Physica D {\bf 77} (1994), 473. 

\bibitem{benilov} E.S. Benilov, R. Grimshaw, and E.P. Kuznetsova,  
 The generation of radiating waves in a singularly-perturbed KdV 
equation, Physica D {\bf 69} (1993), 270. 

\bibitem{joshi} R. Grimshaw and N. Joshi,  Weakly 
nonlocal solitary waves in a singularly perturbed 
Korteweg--de Vries equation, SIAM J. Appl. Math. 
{\bf 55} (1995), 124-135. 

\bibitem{karpman} V.I. Karpman and E.M. Maslov, 
 Perturbation theory for solitons. Sov. Phys. JETP 46 (1978), 281. 

\bibitem{kaupnewell} D.J. Kaup and A.C. Newell,  Solitons 
as particles, oscillators, and in slowly changing media: 
a singular perturbation theory. Proc. R. Soc. Lond. A 
{\bf 361} (1978), 413-446. 

\bibitem{boydbook} J.P. Boyd, {\em Weakly nonlinear solitary waves 
and beyond-all-orders asymptotics}, Kluwer Academic Publishers, 
Boston, 1998. 

\bibitem{craig} W. Craig and M.D. Groves, 
 Hamiltonian long-wave approximations to the 
water-wave problem.'' Wave Motion {\bf 19} (1994), 367. 

\bibitem{milewsky} P.A. Milewsky and E.G. Tabak,  A pseudospectral
procedure for the solution of nonlinear wave equations with examples
from free-surface flows, SIAM. J. Sci. Comp. 21 (1999), 1102. 


\end{thebibliography}
\end{document}